\documentclass[journal]{IEEEtran}

\usepackage{amsmath,graphicx}
\usepackage{amssymb}
\usepackage{array}
\usepackage{amsthm}
\usepackage{cases}

\newtheorem{Theorem}{Theorem}[section]
\newtheorem{Lemma}[Theorem]{Lemma}
\newtheorem{Corollary}[Theorem]{Corollary}
\newtheorem{Remark}[Theorem]{Remark}
\newtheorem{Proposition}[Theorem]{Proposition}
\newtheorem{Definition}[Theorem]{Definition}

\newcounter{claim_nb}[Theorem]
\newcommand{\liuhao}{\fontsize{7.875pt}{\baselineskip}\selectfont}

\begin{document}

\title{New RIC Bounds via $\ell_q$-minimization with $0<q\leq1$ in Compressed Sensing}

\author{Shenglong Zhou$^\dag$, Lingchen Kong$^\dag$, Ziyan Luo$^\ddag$, Naihua Xiu$^\dag$
\thanks{July 28, 2013. $\dag$ Department of Applied Mathematics, Beijing Jiaotong University, Beijing 100044, P. R. China; $\ddag$ State Key Laboratory of Rail Traffic Control and Safety, Beijing Jiaotong University, Beijing 100044, P. R. China (e-mail: longnan\_zsl@163.com, konglchen@126.com, zyluo@bjtu.edu.cn, nhxiu@bjtu.edu.cn).}}

\maketitle

\begin{abstract}
The restricted isometry constants (RICs) play an important role in exact recovery theory of sparse signals via  $\ell_q(0<q\leq1)$ relaxations in compressed sensing. Recently, Cai and Zhang \cite{CZ5} have achieved a sharp bound $\delta_{tk}<\sqrt{1-1/t}$ for $t\geq\frac{4}{3}$ to guarantee the exact recovery of $k$ sparse signals through the $\ell_1$ minimization. This paper aims to establish new RICs bounds via $\ell_q(0<q\leq1)$ relaxation. Based on a key inequality on $\ell_q$  norm, we show that (i) the exact recovery can be succeeded via $\ell_{1/2}$ and $\ell_{1}$ minimizations if $\delta_{tk}<\sqrt{1-1/t}$ for any $t>1$, (ii)several sufficient conditions can be derived, such as for any $q\in(0,\frac{1}{2})$, $\delta_{2k}<0.5547$ when $k\geq2$,  for any  $q\in(\frac{1}{2},1)$, $\delta_{2k}<0.6782$ when $k\geq1$, (iii) the bound on  $\delta_{k}$ is given as well for any $0<q\leq1$, especially for $q=\frac{1}{2},1$, we obtain $\delta_{k}<\frac{1}{3}$ when $k(\geq2)$ is even or $\delta_{k}<0.3203$  when $k(\geq3)$ is odd.
\end{abstract}


\section{Introduction}

\IEEEPARstart{T}{he} concept of compressed sensing (CS) was  initiated  by Donoho \cite{D8}, Cand$\grave{\textmd{e}}$s, Romberg and Tao \cite{CRT9} and Cand$\grave{\textmd{e}}$s and Tao \cite{CT1} with the involved essential idea--recovering some original $n$-dimensional but sparse signal$\setminus$image from linear measurement with dimension far fewer than $n$. Large numbers of researchers, including applied mathematicians, computer scientists and engineers, have paid their attention to this area owing to its wide applications in signal processing, communications, astronomy, biology, medicine, seismology and so on, see, e.g., survey papers \cite{BDE12,R14} and a monograph \cite{EK13}.

To recover a sparse solution $x\in \mathbb{R}^n$ of the underdetermined system of the form $\Phi x = y$, where $y\in \mathbb{R}^m$ is the available measurement and $\Phi \in \mathbb{R}^{m\times n}$ is a known measurement matrix (with $m \ll n$ ), the underlying model is the following $\ell_0$ \emph{minimization}:
\begin{eqnarray}\label{l0} \textup{min} ~\|x\|_0,~~\textup{s.t.}~\Phi x=y,\end{eqnarray}
 where  $\|x\|_0$ is $\ell_0$-norm of the vector $x\in\mathbb{R}^n$, i.e., the number of nonzero
entries in $x$ (this is not a true norm, as $\|\cdot\|_0$ is not
positive homogeneous). However (\ref{l0}) is combinatorial and computationally intractable.

One natural approach is to solve (\ref{l0})  via convex \emph{$\ell_1$ minimization}:
\begin{eqnarray}\label{l1} \textup{min} ~\|x\|_1,~~\textup{s.t.}~\Phi x=y.\end{eqnarray}

The other way is to relax (\ref{l0}) through the nonconvex \emph{$\ell_q(0<q<1)$  minimization}:
\begin{eqnarray}\label{lq} \textup{min} ~\|x\|_q^{q},~~\textup{s.t.}~\Phi x=y,\end{eqnarray}
where $\|x\|_q^{q}=\sum_j|x_j|^q$. Motivated by the fact $\lim\limits_{q\rightarrow0^+}\|x\|_q^{q}=\|x\|_0$, it is shown that there are several advantages of using this approach to recover the sparse signal \cite{LW}. This model for recovering the sparse solution is widely considered, see \cite{C4,CGWY,CXY,DG7,F11,FL10,GN,LW,XS}.

One of the most popular conditions for exact sparse recovery via  $\ell_1$ or $\ell_q$ minimization is related to the \emph{Restricted Isometry Property} (RIP) introduced by Cand$\grave{\textmd{e}}$s and Tao~\cite{CT1}, which was recalled as follows.
\begin{Definition} For $k \in\{ 1, 2,\cdots,n\}$, the restricted isometry constant is the smallest positive number $\delta_k$ such that
\begin{eqnarray}\label{derta} (1-\delta_k)\|x\|_2^2\leq\|\Phi x\|_2^2 \leq(1+\delta_k)\|x\|_2^2\end{eqnarray}
holds for all $k$-sparse vector $x\in\mathbb{R}^n$, i.e., $\|x\|_0\leq k$.
\end{Definition}
It is known that $\delta_{k}$ has the monotone property for $k$ (see, e.g., \cite{CWX2,CWX4}), i.e.,
  \begin{eqnarray}\label{6}\delta_{k_1}\leq \delta_{k_2},~~\textrm{if}~~k_1\leq k_2\leq n.\end{eqnarray}

Current upper bounds on the restricted isometry constants (RICs) via $\ell_q(0<q<1)$ minimization for exact signal recovery were emerged in many studies \cite{C4,DG7,F11,FL10,GN,LW,XS}, such as $\delta_{2k}<0.4531$ for any $q \in (0, 1]$ in \cite{FL10}, $\delta_{2k}<0.4531$ for any $q \in (0, q_0]$ with some $q_0\in (0, 1]$ in \cite{LW} and $\delta_{2k}<0.5$ for any $q \in (0, 0.9181]$ in \cite{XS}. Comparing with those RIC bounds, Cai and Zhang \cite{CZ5} recently have given a sharp bound $\delta_{2k}<\frac{\sqrt{2}}{2}$ via $\ell_1$ minimization.

Motivated by results above, we make our concentrations on improving RIC bounds via $\ell_q$ relaxation with $0<q\leq1$. The main contributions of this paper are the following three aspects:

 (i) If the restricted isometry constant of $\Phi$ satisfies $\delta_{tk}<\sqrt{(t-1)/t}$ for $t>1$, which implies $\delta_{2k}<\frac{\sqrt{2}}{2}$, then exact recovery can be succeeded via $\ell_{\frac{1}{2}}$ and $\ell_{1}$  minimizations.

 (ii) For any $k\geq1$, the bound for $\delta_{2k}$ is an nondecreasing function on $q\in(0,\frac{1}{2})$ and $q\in(\frac{1}{2},1)$. Moreover, several sufficient conditions are derived, such as for any $q\in(0,\frac{1}{2})$, $\delta_{2k}<0.5547$ when $k\geq2$, for any  $q\in(\frac{1}{2},1)$, $\delta_{2k}<0.6782$ when $k\geq1$. The detailed can be seen in Tab. 2 of the Section III, which are all better bounds than current ones  in terms of  $\ell_q (0<q<1)$ minimization.

 (iii) The bound on  $\delta_{k}$ is given as well for any $0<q\leq1$. Especially for $q=\frac{1}{2},1$, we obtain $\delta_{k}<\frac{1}{3}$ when $k$ is even or $\delta_{k}<0.3203$  when $k(\geq3)$ is odd.

The organization of this paper is as follows. In the next section, we establish several key lemmas. Our main results on $\delta_{tk}$ with $t>1$ and  $\delta_{k}$ will be presented in  Sections III and IV respectively. We make some concluding remarks in Section V and give the proofs of all lemmas and theorems in the last section.

\section{Key Lemmas}
This section will propose several technical lemmas, which play an important role in the sequel analysis. We begin with recalling the lemma of the sparse representation of a polytope stated by  Cai and Zhang \cite{CZ5}. Here, we define $\|x\|_{\infty}:=\textmd{max}_i~\{|x_i|\}$ and $\|x\|_{-\infty}:=\textmd{min}_i~\{|x_i|\}$ (In fact, $\textit{l}_{-\infty}$ is not a norm since the triangle inequality fails).
\begin{Lemma}\label{lemma1}
For a positive number $\alpha$ and a positive integer $s$, define the polytope $T(\alpha, s)\subset \mathbb{R}^n$ by
$$T(\alpha,s) = \left\{v\in\mathbb{R}^n \large ~|~ \|v\|_\infty \leq\alpha, \|v\|_1\leq s\alpha\right\}.$$
For any $v\in\mathbb{R}^n$, define the set $U(\alpha, s, v)\subset \mathbb{R}^n$ of sparse vectors by
 \vspace{-6mm}
\begin{flushright}\begin{eqnarray}U(\alpha,s, v) =\{u\in\mathbb{R}^n \large~|~supp(u)\subseteq supp(v),\|u\|_0\leq s,\nonumber\\
\|u\|_1 = \|v\|_1,\|u\|_\infty \leq \alpha\}.\nonumber\end{eqnarray}\end{flushright}
Then $v\in T(\alpha,s)$ if and only if $v$ is in the convex hull of $U(\alpha, s, v)$. In particular, any $v\in T(\alpha,s)$ can be expressed as $v=\sum_{i=1}^{N}\lambda_iu_i,$ where~$$0\leq\lambda_i\leq1,\sum_{i=1}^{N}\lambda_i=1,u_i\in U(\alpha, s, v),i=1,2,\cdots,N.$$
\end{Lemma}

Next we establish an interesting and important inequality in the following lemma, which gives a sharpened estimation of $\ell_1$ with $\ell_0,\ell_q, \ell_\infty$ and $\ell_{-\infty}$.

\begin{Lemma}\label{lemma2}For $q\in(0,1]$ and $x\in\mathbb{R}^n$, we have
\begin{eqnarray}\label{l1q}
\|x\|_1\leq\frac{\|x\|_q}{n^{1/q-1}}+p_qn(\|x\|_\infty-\|x\|_{-\infty}),\end{eqnarray}
where
\begin{eqnarray}\label{pq}p_q:=q^{\frac{q}{1-q}}-q^{\frac{1}{1-q}}.\end{eqnarray}
Moreover, $p_q$ is a nonincreasing and convex  function of $q\in[0,1]$ with
$$p_0:=\lim_{q\rightarrow0^+}p_q=1~and~p_1:=\lim_{q\rightarrow1^-}p_q=0.$$
\end{Lemma}

\makeatletter
          \def\@captype{figure}
          \makeatother
          \begin{center}
          \includegraphics[width=7.5cm,height=4.3cm]{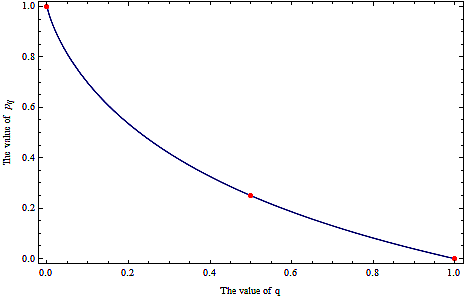}
          \end{center}
           \vspace{-5mm}
\caption{Plot of $p_q\in[0,1]$ as a function of $q\in[0,1]$, and $p_{\frac{1}{2}}=\frac{1}{4}$.\label{fig1}}

\begin{Remark} Actually, we can substitute $n$ with $\|x\|_0$ in inequality (\ref{l1q}), which leads to
\begin{eqnarray}\label{l1q0}
\|x\|_1\leq\frac{\|x\|_q}{\|x\|_0^{1/q-1}}+p_q\|x\|_0(\|x\|_\infty-\|x\|_{-\infty}).\end{eqnarray}\end{Remark}
Moreover, combining with the H$\ddot{o}$lder Inequality and $\left(\ref{l1q0}\right)$, we have
\begin{Proposition}For $q\in(0,1]$ and $x\in\mathbb{R}^n$, we have
\begin{eqnarray}\label{1q}\|x\|_0^{1-\frac{1}{q}}\|x\|_q\leq\|x\|_1\leq\left(\|x\|_0^{1-\frac{1}{q}}+p_q\|x\|_0\right)\|x\|_q.\end{eqnarray}\end{Proposition}
 \noindent Here, $\left(\ref{1q}\right)$ is an interesting inequality. Although $\left(\ref{1q}\right)$ will not be applied in our proof, it manifests the relationship between $\ell_1$ and $\ell_q$ norm.

In order to analyze a sequent useful function more clearly, we first observe the function $q^{\frac{q}{q-1}}$ of $q\in(0,1)$, whose figure is plotted below.
\vspace{2mm}
\makeatletter
          \def\@captype{figure}
          \makeatother
          \begin{center}
          \includegraphics[width=7.5cm,height=4.3cm]{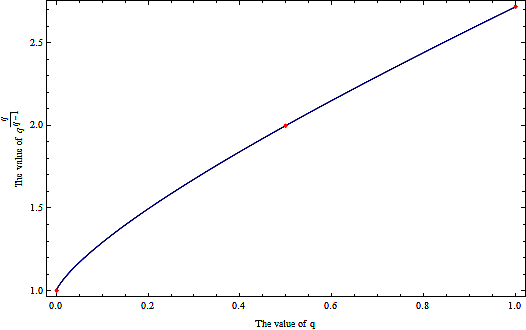}
          \end{center}
           \vspace{-5mm}
\caption{Plot of $q^{\frac{q}{q-1}}$ as a function of $q\in[0,1]$.\label{fig2}}
\vspace{2mm}
 \noindent It is easy to check that
\begin{eqnarray}\label{q01}
  \lim_{q\rightarrow0^+}q^{\frac{q}{q-1}}=1,~~\lim_{q\rightarrow1^-}q^{\frac{q}{q-1}}=e.\end{eqnarray}
  So $q^{\frac{q}{q-1}}$ can be defined as a function of $q$  on $[0,1]$, and it is a nondecreasing function.

In addition, for any given integer $k\geq1$, it is trivial that if $q^{\frac{q}{q-1}}$ is an integer, then $q^{\frac{q}{q-1}}k$ apparently is an integer as well for instance $q=1/2$. However, the integrity of $q^{\frac{q}{q-1}}$ is  not necessary to ensure the integrity of $q^{\frac{q}{q-1}}k$, such as $q=2/3$ and $k = 4$.

Based on analysis above, we now define a real valued function $g(q,k):(0,1)\times\{1,2,3,\cdots\}\rightarrow\mathbb{R}$ by
 \begin{eqnarray}\label{g}&&g(q,k):=\lceil q^{\frac{q}{q-1}}k\rceil^{1-1/q}k^{1/q}+p_q\lceil q^{\frac{q}{q-1}}k\rceil,\nonumber\\
 &&~~~~~~~~~~~~~~~~~~~q\in(0,1),~k\in\{1,2,3,\cdots\},\end{eqnarray}
 where $p_q$ is defined as in (\ref{pq}) and $\lceil a \rceil$ denotes the smallest integer that is no less than $a$.

\begin{Lemma}\label{lemma3} Let $g(q,k)$ be defined as in $\left(\ref{g}\right)$. Then $g(q,k)=k$ when $q^{\frac{q}{q-1}}k$ is an integer and otherwise $g(q,k)\leq k+p_q$. Moreover,
\begin{eqnarray}&&g(0,k):=\lim\limits_{q\rightarrow0^+}g(q,k)=k+1,\nonumber\\
&&g(1,k):=\lim\limits_{q\rightarrow1^-}g(q,k)=k.\nonumber\end{eqnarray}\end{Lemma}
\noindent   Therefore, $g(q,k)$  can be regarded as a function of $q$ on $[0,1]$, and the image of $g(q,k)$ with the special case $k=1$, where $g(0,1)=2,g(\frac{1}{2},1)=1,g(1,1)=1$, is plotted in Fig.\ref{fig3}.
\makeatletter
          \def\@captype{figure}
          \makeatother
          \begin{center}
          \includegraphics[width=7.5cm,height=4.5cm]{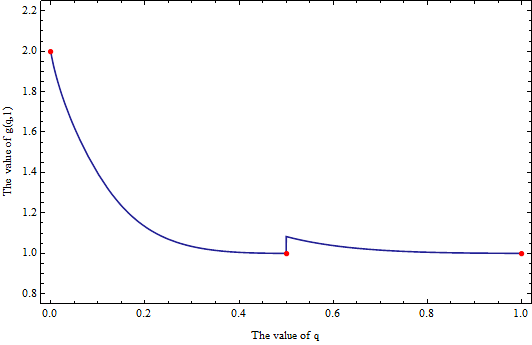}
          \end{center}
           \vspace{-5mm}
\caption{Plot of $g(q,1)$ as a function of $q\in[0,1]$.\label{fig3}}
\vspace{2mm}
Another two useful functions are introduced and analyzed in the following lemma, which will ease sequent analysis of our main results.

\begin{Lemma}\label{lemma4}For $t>1$ and $\theta\geq0, \rho\geq0$, we define
\begin{eqnarray}\label{mu}\mu(t,\theta)&:=&\frac{\sqrt{(t+\theta-1)(t-1)}+1-t}{\theta},\\
\label{derta}\gamma(\rho,\theta)&:=&\frac{\rho-\rho^2}{\frac{1}{2}-\rho+\rho^2(1+\frac{\theta}{2(t-1)})}.\end{eqnarray}
Then $\gamma\left(\mu\left(t,\theta\right),\theta\right)$ is a nonincreasing  function on $\theta$ when $t$ is fixed while a nondecreasing function on $t$ when $\theta$ is fixed.
\end{Lemma}

\section{ Main Results on $\delta_{tk}$ with $t>1$}
Now we give our main results on $\delta_{tk}$ with $t>1$:
\begin{Theorem}\label{theorem}For any $q\in(0,1]$, if \begin{eqnarray}\label{dertagk}\delta_{g(q,k)(t-1)+k}< \gamma\left(\mu\left(t,\frac{g(q,k)}{k}\right),\frac{g(q,k)}{k}\right)\end{eqnarray} holds for some $t>1$, then each $k$-sparse minimizer of the $\ell_q$ minimization $(\ref{lq})$ is the sparse solution of $(\ref{l0})$. Furthermore, setting $t=1+\frac{(\tau-1)k}{g(q,k)}$ with $\tau>1$, then the sufficient condition $\left(\ref{dertagk}\right)$ of exact signal recovery can be reformulated as
 \begin{eqnarray}\label{taok0}\delta_{\tau k}< \gamma\left(\mu\left(1+\frac{(\tau-1)k}{g(q,k)},\frac{g(q,k)}{k}\right),\frac{g(q,k)}{k}\right).\end{eqnarray}
\end{Theorem}
From Lemma \ref{lemma3}, when $q=1$ or $q^{\frac{q}{q-1}}k$ is an integer (such as $q=\frac{1}{2}$), it follows that $g(q,k)=k$. Associating with (\ref{dertagk}) in Theorem \ref{theorem}, we have $\delta_{tk}=\delta_{g(q,k)(t-1)+k}< \gamma\left(\mu\left(t,1\right),1\right)=\sqrt{\frac{t-1}{t}}$. Therefore, a corollary can be elicited as below.

\begin{Corollary}\label{prop1}
For $q=1$ or $q\in(0,1)$ such that $q^{\frac{q}{q-1}}k$ is an integer, if $\delta_{tk}<\sqrt{\frac{t-1}{t}}$ holds with some $t>1$ and $k\geq1$, then each $k$-sparse minimizer of the $\ell_q$ minimization  $\left(\ref{lq}\right)$ is the sparse solution of $\left(\ref{l0}\right)$.
\end{Corollary}
\noindent In particular, taking $t=2,3,4$, we obtain $\delta_{2k}<\frac{\sqrt{2}}{2}\approx0.7071$, $\delta_{3k}<0.8164,~\delta_{4k}<0.8660$ respectively. It is worth mentioning that $\delta_{tk}<\sqrt{\frac{t-1}{t}}$ is the sharp bound for $\ell_1$ minimization which has been proved by Cai and Zhang \cite{CZ5}. Because exact recovery can fail for any $q\in(0,1]$ if the bound of $\delta_{2k}$ is no less than $\frac{\sqrt{2}}{2}$ (see \cite{DG7}), $\delta_{2k}<\frac{\sqrt{2}}{2}$ is also the sharp bound for $\ell_{\frac{1}{2}}$ minimization.

Actually, besides $q=\frac{1}{2}$, $k\geq1$, there are several other $(q,k)$s satisfying that $q^{\frac{q}{q-1}}k$ are integers, for instance $(0.2025,2)$, $(\frac{2}{3},4)$. Thus $\delta_{tk}<\sqrt{\frac{t-1}{t}}$ is also a sharp RIC bound for such $(q,k)$s.

\begin{Remark}(i) For any $k\geq1$, we can check $$g(q,1)\geq\frac{g(q,k)}{k}.$$ Then from Lemma \ref{lemma4} and $\left(\ref{taok0}\right)$ in Theorem \ref{theorem}, for $k\geq1$ and any $q\in(0,1]$, it yields that
 \begin{eqnarray}\label{taokk}\delta_{\tau k}<\gamma\left(\mu\left(1+\frac{\tau-1}{g(q,1)},g(q,1)\right),g(q,1)\right),\end{eqnarray}
 whose figure (with $\tau=2$) is plotted as follows.
 \vspace{2mm}
\makeatletter
          \def\@captype{figure}
          \makeatother
          \begin{center}
          \includegraphics[width=7.5cm,height=4.6cm]{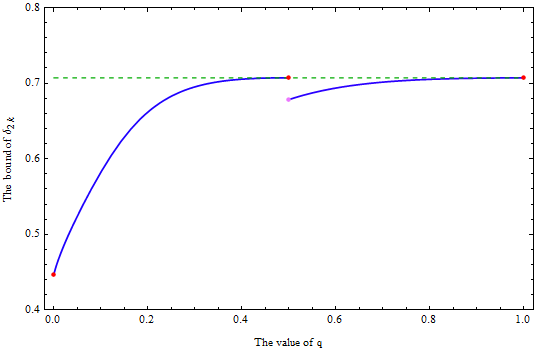}
\vspace{-2mm}
         \caption{Plot of bounds on $\delta_{2k}$ as a function of $q\in(0,1]$ when $k\geq1$.\label{fig2k}}
          \end{center}

(ii) Moreover, under some assumptions $k\geq k_0(k_0=1,2,3,\cdots)$, since for $q\in(\frac{1}{2},1]$
$$\lim_{q\rightarrow\frac{1}{2}^{+}}\frac{g(q,k_0)}{k_0}\geq\max \{\lim_{q\rightarrow\frac{1}{2}^{+}}\frac{g(q,k)}{k},~\frac{g(q,k_0)}{k_0} \}$$
and for $q\in(0,\frac{1}{2}]$
$$\lim_{q\rightarrow0^{+}}\frac{g(q,k_0)}{k_0}\geq\max \{\lim_{q\rightarrow0^{+}}\frac{g(q,k)}{k},~\frac{g(q,k_0)}{k_0}\}.$$
 Then from Lemma \ref{lemma4}, we have Tab. 2 by calculating limits for cases $q\rightarrow0^+$ and $q\rightarrow\frac{1}{2}^+$  of the right-hand side of $(\ref{taok0})$ with  $k = k_0$.
\makeatletter
          \def\@captype{figure}
          \makeatother
          \begin{center}
          \includegraphics[width=8.2cm,height=3.7cm]{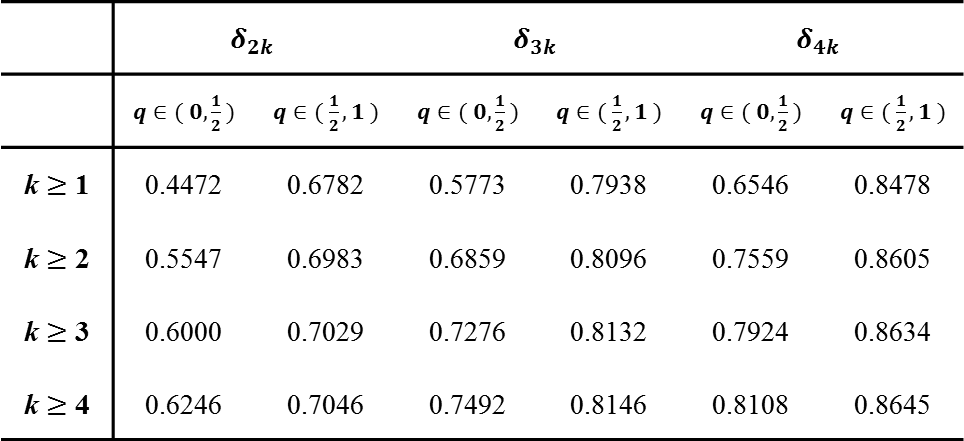}
\end{center}
          \begin{flushleft}
          \liuhao{\emph{Tab. 2: Bounds on $\delta_{2k},\delta_{3k},\delta_{4k}$ for any $q\in(0,\frac{1}{2})$ and $q\in(\frac{1}{2},1)$.}}
          \end{flushleft}

 \end{Remark}

\section{ Main Results on $\delta_{k}$ }
In this section, we state the bound on $\delta_{k}$ for any $q\in(0,1]$ in the following results:
\begin{Theorem}\label{theorem1} For any $q\in(0,1]$, if
\begin{numcases}{\delta_{k}<}
~~~~\frac{1}{1+2\lceil g(q,k)\rceil/k},~~~~~~\text{for even number}~ k\geq2,\nonumber\\
\frac{1}{1+2\lceil g(q,k)\rceil /\sqrt{k^{2}-1}},~~ \text{for odd number}~ k\geq3,\nonumber\end{numcases}
holds, then each $k$-sparse minimizer of the $\ell_q$ minimization $(\ref{lq})$ is the sparse solution of $(\ref{l0})$.
\end{Theorem}
Particularly, for the case $q=1$ or $q^{\frac{q}{q-1}}k$ to be an integer (such as $q=\frac{1}{2}$ ), we have the corollary below by applying Lemma \ref{lemma3}.
\begin{Corollary}\label{prop3}
For $q=1$ or $q\in(0,1)$ such that $q^{\frac{q}{q-1}}k$ is an integer, if
\begin{numcases}{\delta_{k}<}
~~~~~~~~1/3,~~~~~~~~~~\text{for even number}~ k\geq2,\nonumber\\
\frac{1}{1+2k /\sqrt{k^{2}-1}},~~ \text{for odd number}~ k\geq3,\nonumber\end{numcases} hold, then each $k$-sparse minimizer of the $\ell_q$ minimization $(\ref{lq})$ is the sparse solution of $(\ref{l0})$.
\end{Corollary}
Taking $q=\frac{1}{2}, 1$, then $g(q,k) = k$ from Lemma \ref{lemma3}, which produces the bound $\delta_{k}<\frac{1}{3}$ if $k\geq2$ is even. Meanwhile $\delta_{k}<\frac{1}{3}$ for $k\geq2$ is the sharp bound for $\ell_1$ minimization that has been gotten by Cai and Zhang \cite{CZ}. From Theorem \ref{theorem1} and Corollary \ref{prop3}, we list the following table.

 \vspace{-0.5mm}
\begin{figure}[h]
\centering\includegraphics[width=8cm,height=2.7cm]{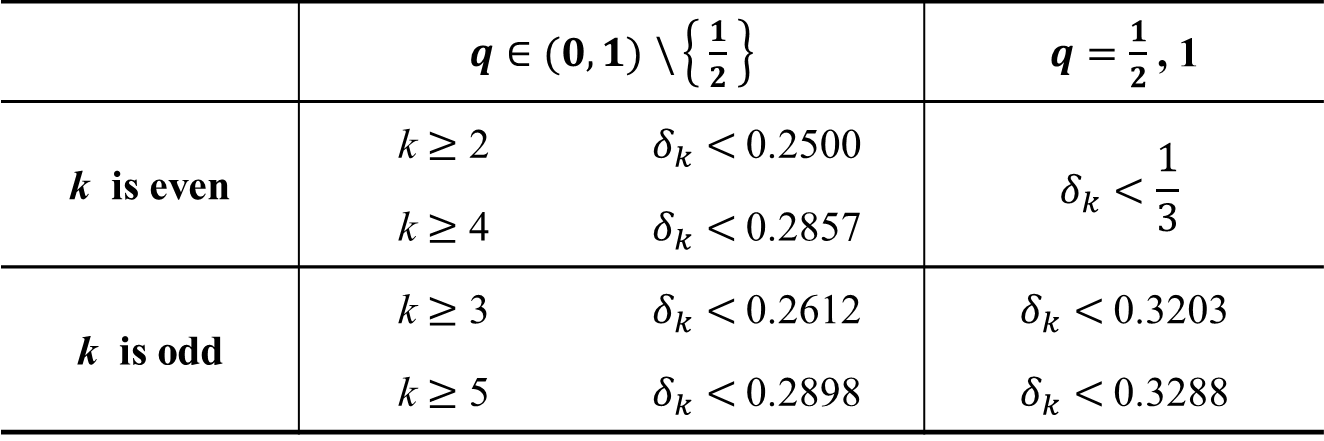}
\begin{flushleft}
 \liuhao{Tab. 3: Upper bounds on $\delta_k$ for different $q$.}
\end{flushleft}
\end{figure}

\vspace{-4mm}
\section{Concluding Remarks}
In this  paper,  we have generalized the upper bounds for RICs from $\ell_1$ minimization to $\ell_q(0<q\leq1)$ minimization, and established new RIC bounds through $\ell_q$ minimization with $q\in(0,1]$ for exact sparse recovery. An interesting issue which deserves future research would be: how to improve these new bounds for some $q\in(0,1]$ when $q^{\frac{q}{q-1}}k$ is not an integer.

\section{Proofs}

\noindent\textbf{Proof of Lemma \ref{lemma2}}\\
Stimulated by the approach in \cite{XS}, without loss of generality, we only need to prove the case $x \in\Omega:= \{(x_1,x_2,\cdots,x_n)\neq0~|
~x_1\geq x_2\geq\cdots\geq x_n\geq 0 \}$ due to the symmetry of components $|x_1|,|x_2|,\cdots,|x_n|$. Clearly, $x_1\neq0$. Notice that if the inequality (\ref{l1q}) holds for any $(1,x_2,\cdots,x_n) \in \Omega$, then we can immediately generalize the conclusion to all $x \in \Omega$ through substituting $x/x_1, x\in \Omega$ into (\ref{l1q}) and eliminating the common factor $1/x_1$. Henceforth, it remains to show
\begin{eqnarray}\label{l1q1}
\|x\|_1\leq\frac{\|x\|_q}{n^{1/q-1}}+p_qn(1-x_n),\end{eqnarray}
with $x\in \{(1,x_2,\cdots,x_n)~|~1\geq x_2\geq\cdots\geq x_n\geq 0 \}$, where $p_q$ is a function of $q$ specified in (\ref{pq}).

First, for any given $q\in(0,1]$ define that
$$f(x):=\|x\|_1-n^{1 -1/q}\|x\|_q.$$
It is easy to verify that $f(x)$ is a convex function on $\mathbb{R}^n_+$. Since the maximum of a convex function always arrives on the boundary, we have
\begin{eqnarray}
h(x_n):&=&\max_{1\geq x_2\geq x_3\geq\cdots\geq x_n}~f(1, x_2, x_3,\cdots,x_n)\nonumber\\
&=&f(1,\cdots,1, x_n,\cdots,x_n),~~x_n\in[0,1]\nonumber\end{eqnarray}
Letting the distribution of $1$ appear for $r$ times ($1 \leq r \leq n$) in the maximum solution of $f$, we have
$$h(x_n)=r(1-x_n)+nx_n-\frac{\left(r(1-x_n^q)+nx_n^q\right)^{1/q}}{n^{1/q-1}}.$$
By the convexity of $h$ and $h(1)=0$, it follows that
$$h(x_n)\leq(1-x_n)h(0)+x_nh(1)=(1-x_n)h(0).$$
Then it holds that
\begin{eqnarray}
f(x)&\leq &h(x_n)\leq(1-x_n)h(0)\nonumber\\
&=&(1-x_n)(r-n^{1-1/q}r^{1/q})\nonumber\\
&\leq&(1-x_n)\max_{r\in\{1,2,\cdots,n\}}\{r-n^{1-1/q}r^{1/q}\}\nonumber\\
&\leq&(1-x_n)\max_{0<r_1\leq n}\{r_1-n^{1-1/q}r_1^{1/q}\}\nonumber\\
&=&(1-x_n)p_qn,\nonumber\end{eqnarray}
where $p_q$ is defined as (\ref{pq}) and the last equality holds when $r_1=q^{\frac{q}{1-q}}n\in (0,n]$ for any $q\in (0,1]$.

By computing the first and second order partial derivatives of $p_q$ on $q$, it is easy to verify that $p_q$ is a nonincreasing convex function of $q\in(0,1]$ and
$$\lim_{q\rightarrow0^+}p_q=1~\textmd{and}~\lim_{q\rightarrow1^-}p_q=0.$$
Thus the proof is completed. \qed

\vspace{2mm}
\noindent\textbf{Proof of Lemma \ref{lemma3}}\\
If $q^{\frac{q}{q-1}}k$ is an integer, then
\begin{eqnarray}g(q,k)&=&(q^{\frac{q}{q-1}}k)^{1-1/q}k^{1/q}+p_q( q^{\frac{q}{q-1}}k)\nonumber\\
&=&qk^{1-1/q}k^{1/q}+(q^{\frac{q}{1-q}}-q^{\frac{1}{1-q}})( q^{\frac{q}{q-1}}k)\nonumber\\
&=&qk+(1-q)k=k.\nonumber\end{eqnarray}
If $q^{\frac{q}{q-1}}k$ is not an integer, then
\begin{eqnarray}g(q,k)&\leq&(q^{\frac{q}{q-1}}k)^{1-1/q}k^{1/q}+p_q( q^{\frac{q}{q-1}}k+1)\nonumber\\
&=&qk^{1-1/q}k^{1/q}+(q^{\frac{q}{1-q}}-q^{\frac{1}{1-q}})( q^{\frac{q}{q-1}}k+1)\nonumber\\
&=&qk+(1-q)k+p_q=k+p_q.\nonumber\end{eqnarray}
Due to  $\lim_{q\rightarrow1^-}q^{\frac{q}{q-1}}=e$ and $\lim_{q\rightarrow1^-}p_q=0$ , we have
\begin{eqnarray}g(1,k):&=&\lim_{q\rightarrow1^-}g(q,k)\nonumber\\
&=&\lim_{q\rightarrow1^-}\left\{\lceil q^{\frac{q}{q-1}}k\rceil^{1-1/q}k^{1/q}+p_q\lceil q^{\frac{q}{q-1}}k\rceil\right\}\nonumber\\
&=&k+0=k.\nonumber\end{eqnarray}
Now we prove the remaining part $\lim_{q\rightarrow0^+}g(q,k)=k+1$. Since $\lim_{q\rightarrow0^+}q^{\frac{q}{q-1}}=1$ and $q^{\frac{q}{q-1}}\in(1,e]$ is a nondecreasing function on $q\in(0,1]$, for any fixed $k$, we can set
$q^{\frac{q}{q-1}}=1+\varepsilon(q)$ with sufficient small $0<\varepsilon(q)<\frac{1}{k}$. Thus
$$\lceil q^{\frac{q}{q-1}}k\rceil=\lceil (1+\varepsilon(q))k\rceil=k+1,~~\text{as}~ q(\neq0)\rightarrow0^+ ,$$
It follows readily that
\begin{eqnarray}g(0,k):&=&\lim_{q\rightarrow0^+}g(q,k)\nonumber\\
&=&\lim_{q\rightarrow0^+}\left\{\lceil q^{\frac{q}{q-1}}k\rceil^{1-1/q}k^{1/q}+p_q\lceil q^{\frac{q}{q-1}}k\rceil\right\}\nonumber\\
&=&\lim_{q\rightarrow0^+}\left\{( k+1)^{1-1/q}k^{1/q}+p_q(k+1)\right\}\nonumber\\
&=&\lim_{q\rightarrow0^+}\left\{( k+1)\left(\frac{k}{k+1}\right)^{1/q}+p_q(k+1)\right\}\nonumber\\
&=&0+k+1=k+1.\nonumber\end{eqnarray}
The whole proof is finished.\qed

\vspace{2mm}
\noindent\textbf{Proof of Lemma \ref{lemma4}}\\
We verify $\gamma\left(\mu\left(t,\theta\right),\theta\right)$ is a nonincreasing function on $\theta\geq0$ and a nondecreasing function on $t>1$. By directly computing the first order partial  derivative of $\gamma\left(\mu\left(t,\theta\right),\theta\right)$ on $\theta\geq0$, it yields
$$\frac{\partial}{\partial\theta}\gamma\left(\mu\left(t,\theta\right),\theta\right)=\frac{-\sqrt{(t+\theta-1)(t-1)}}{2(t+\theta-1)^2}\leq0.$$
Likewise, by computing the first order partial derivative of $\gamma\left(\mu\left(t,\theta\right),\theta\right)$ on $t>1$, we have
$$\frac{\partial}{\partial t}\gamma\left(\mu\left(t,\theta\right),\theta\right)=\frac{\theta}{2\sqrt{(t-1)(t+\theta-1)^3}}\geq0.$$
Then the desired conclusions hold immediately.\qed

Before proving Theorem \ref{theorem}, we introduce hereafter several notations. For $h\in\mathbb{R}^n$, we denote hereafter $h_T$ the vector equal to $h$ on an index set $T$ and zero elsewhere. Especially, we denote $h_{max(k)}$ as $h$ with all but the largest $k$ entries in absolute value set to zero, and $h_{-max(k)} := h -h_{max(k)}$.

\vspace{2mm}
\noindent\textbf{Proof of Theorem \ref{theorem}}\\
 The approach of this proof is similar as \cite{CZ5}. First we consider the case that $g(k,q)(t-1)$ is an integer. By the Null Space Property \cite{LW} in $\ell_q$ minimization case, we only need to check for all $h \in \mathcal{N}(\Phi)\setminus\{0\}$,
$$\|h_{max(k)}\|_{q}^{q}<\|h_{-max(k)} \|_{q}^{q}.$$
 Suppose on the contrary that there exists $h \in \mathcal{N}(\Phi)\setminus\{0\}$, such that $\|h_{max(k)}\|_{q}^{q}\geq\|h_{-max(k)} \|_{q}^{q}$. Set $\alpha=k^{-1/q}\|h_{max(k)}\|_{q}$ and  decompose $h_{-max(k)}$ into a sum of vectors $h_{T_1},h_{T_2},\ldots $, where $T_1$ corresponds to the locations of the $\lceil q^{\frac{q}{q-1}}k\rceil$ largest coefficients of  $h_{-max(k)}$  ; $T_2$ to the locations of the $\lceil q^{\frac{q}{q-1}}k\rceil$ largest coefficients of $h_{-max(k)T_1^C}$,  and so on. That is
  \begin{eqnarray}h_{-max(k)}=h_{T_1} +h_{T_2}+h_{T_3}+ \cdots .\nonumber\end{eqnarray}
Here, the sparsity of $h_{T_j}(j\geq1)$ is at most $\lceil q^{\frac{q}{q-1}}k\rceil$.

Clearly, $k\|h_{-max(k)} \|_{\infty}^{q}\leq\|h_{max(k)} \|_{q}^{q}=k\alpha^{q}$, which generates $\|h_{-max(k)} \|_{\infty}\leq\alpha$. From Lemma \ref{lemma2}, for $j\geq1$,
\begin{eqnarray}
\label{IN1}\|h_{T_j}\|_1&\leq& \lceil q^{q/(q-1)}k\rceil^{1-1/q} \|h_{T_j}\|_q\nonumber\\
&&+p_q\lceil q^{\frac{q}{q-1}}k\rceil(\|h_{T_j}\|_{\infty}-\|h_{T_j}\|_{-\infty}).~~~~~~\end{eqnarray}
Then we sum $\|h_{T_j}\|_1$ for $j\geq1$ to obtain that
\begin{eqnarray}\label{IN2}&&\|h_{-max(k)} \|_1=\sum_{j\geq1}\|h_{T_j}\|_1\nonumber\\
&\leq&\lceil q^{\frac{q}{q-1}}k\rceil^{1-1/q}\sum_{j\geq1}\|h_{T_j}\|_q \nonumber\\
&&+p_q\lceil q^{\frac{q}{q-1}}k\rceil\sum_{j\geq1}\left(\|h_{T_j}\|_{\infty}-\|h_{T_j}\|_{-\infty}\right)\nonumber\\
&\leq&\lceil q^{\frac{q}{q-1}}k\rceil^{\frac{q-1}{q}}(\sum_{j\geq1}\|h_{T_j}\|_q^q)^{1/q} +p_q\lceil q^{\frac{q}{q-1}}k\rceil\|h_{T_1}\|_{\infty}.\nonumber\\
&\leq&\lceil q^{\frac{q}{q-1}}k\rceil^{\frac{q-1}{q}}k^{1/q}\alpha +p_q\lceil q^{\frac{q}{q-1}}k\rceil\alpha=g(q,k)\alpha.\end{eqnarray}

We again divide $h_{-max(k)}$ into two parts, $h_{-max(k)}= h^{(1)} + h^{(2)}$, where
$$h^{(1)} := h\cdot\textbf{1}_{\{i:|h_{-max(k)}(i)|>\frac{\alpha}{t-1}\}},$$
$$h^{(2)} := h\cdot\textbf{1}_{\{i:|h_{-max(k)}(i)|\leq\frac{\alpha}{t-1}\}}.$$
Therefore $h^{(1)}$ is $g(q,k)(t-1)$-sparse as a result of facts that $\|h^{(1)}\|_1\leq\|h_{-max(k)}\|_1\leq g(q,k)\alpha$ and all non-zero entries of $h^{(1)}$ has magnitude larger than $\frac{\alpha}{t-1}$. Let $\|h^{(1)}\|_0 = m$, then
\begin{eqnarray}\|h^{(2)}\|_1&=&\|h_{max(k)}\|_1-\|h^{(1)}\|_1\nonumber\\
&\leq&\left[g(q,k)(t-1)-m\right]\frac{\alpha}{t-1},\\
\|h^{(2)}\|_\infty&\leq&\frac{\alpha}{t-1}.\end{eqnarray}

Applying Lemma \ref{lemma1} with $s =g(q,k)(t-1)-m$, it makes $h^{(2)}$ be expressed as a convex combination of sparse vectors: $h^{(2)} =\sum_{i=1}^{N}\lambda_iu_i$, where $u_{i}$ is $s$-sparse, $\|u_{i}\|_1=\|h^{(2)}\|_1, \|u_{i}\|_\infty\leq\frac{\alpha}{t-1}$. Henceforth,
$$\|u_{i}\|_2\leq\sqrt{g(q,k)(t-1)-m}\|u_{i}\|_\infty\leq\sqrt{\frac{g(q,k)}{t-1}}\alpha.$$

For any $\mu\geq0$, denoting $\eta_i = h_{max(k)}+ h^{(1)}+\mu u_{i}$, we obtain
\begin{eqnarray}
&&\sum_{j=1}^{N}\lambda_j\eta_j-\frac{1}{2}\eta_i=h_{max(k)}+ h^{(1)}+\mu h^{(2)}-\frac{1}{2}\eta_i~~~~\nonumber\\
\label{a}&=&(\frac{1}{2}-\mu)\left(h_{max(k)}+ h^{(1)}\right)-\frac{1}{2}\mu u_i+\mu h,\end{eqnarray}
where $\eta_i,\sum_{i=1}^{N}\lambda_i\eta_i-\frac{1}{2}\eta_i-\mu h$ are all $\left(g(q,k)(t-1)+k\right)$-sparse vectors from the sparsity of $\|h_{max(k)}\|_0\leq k$,~$\|h^{(1)}\|_0=m$ and $\|u_{i}\|_0\leq s$.

It is easy to check the following identity,
\begin{eqnarray}\label{b}
\sum_{i=1}^{N}\lambda_i\|\Phi(\sum_{j=1}^{N}\lambda_j\eta_j-\frac{1}{2}\eta_i)\|_2^2=\frac{1}{4}\sum_{i=1}^{N}\lambda_i\|\Phi\eta_i\|_2^2.
\end{eqnarray}
Since $\Phi h = 0$, together with (\ref{a}), we have $$\Phi(\sum_{j=1}^{N}\lambda_j\eta_j-\frac{1}{2}\eta_i)=\Phi((\frac{1}{2}-\mu)(h_{max(k)}+ h^{(1)})-\frac{1}{2}\mu u_i).$$
Setting $\mu=\mu\left(t,g(q,k)/k\right)>0$, if (\ref{dertagk}) holds, that is
\begin{eqnarray}\label{bb}\delta:=\delta_{g(q,k)(t-1)+k}<\gamma\left(\mu\left(t,\frac{g(q,k)}{k}\right),\frac{g(q,k)}{k}\right),\end{eqnarray}
then combining (\ref{b}) with (\ref{bb}), we get
\begin{eqnarray}
0&=&\sum_{i=1}^{N}\lambda_i\|\Phi((\frac{1}{2}-\mu)(h_{max(k)}+ h^{(1)})-\frac{1}{2}\mu u_i)\|_2^2\nonumber\\
&&-\frac{1}{4}\sum_{i=1}^{N}\lambda_i\|\Phi\eta_i\|_2^2\nonumber\\
&\leq&(1+\delta)\sum_{i=1}^{N}\lambda_i[(\frac{1}{2}-\mu)^2\|h_{max(k)}+ h^{(1)}\|_2^2+\frac{\mu^{2}}{4}\|u_i\|_2^2]\nonumber\\
&&-\frac{1-\delta}{4}\sum_{i=1}^{N}\lambda_i(\|h_{max(k)}+ h^{(1)}\|_2^2+\mu^{2}\|u_i\|_2^2)\nonumber\\
&=&\sum_{i=1}^{N}\lambda_i[((1+\delta)(\frac{1}{2}-\mu)^2-\frac{1-\delta}{4})\cdot\nonumber\\
&&\|h_{max(k)}+ h^{(1)}\|_2^2+\frac{1}{2}\delta\mu^{2}\|u_i\|_2^2]\nonumber\end{eqnarray}
\begin{eqnarray}&\leq&\sum_{i=1}^{N}\lambda_i\|h_{max(k)}+ h^{(1)}\|_2^2\cdot\nonumber\\
\label{c}&&\left[\mu^2-\mu+\delta(\frac{1}{2}-\mu+(1+\frac{g(q,k)}{2k(t-1)})\mu^{2})\right]\\
&=&\|h_{max(k)}+ h^{(1)}\|_2^2\cdot\nonumber\\
&&\left[\mu^2-\mu+\delta(\frac{1}{2}-\mu+(1+\frac{g(q,k)}{2k(t-1)})\mu^{2})\right]\nonumber\\
&=&\|h_{max(k)}+ h^{(1)}\|_2^2(\frac{1}{2}-\mu+(1+\frac{g(q,k)}{2k(t-1)})\mu^{2})\cdot\nonumber\\
&&\left[\delta-\gamma(\mu\left(t,\frac{g(q,k)}{k}),\frac{g(q,k)}{k}\right)\right]\nonumber\\
&<&0, \nonumber\end{eqnarray}
where the inequality (\ref{c}) is derived from the following facts:
\begin{eqnarray}
\|h_{max(k)}\|_2^2&\geq& k^{1-2/q}\|h_{max(k)}\|_q^2\nonumber\\
\label{hold}&=&k^{1-2/q}(k\alpha^{q})^{2/q}=k\alpha^{2},\\
\|u_{i}\|_2&\leq&\sqrt{\frac{g(q,k)}{t-1}}\alpha\leq\sqrt{\frac{g(q,k)}{k}}\frac{\|h_{max(k)}\|_2}{\sqrt{t-1}}~~~~~~\nonumber\\
&\leq&\sqrt{\frac{g(q,k)}{k}}\frac{\|h_{max(k)}+ h^{(1)}\|_2}{\sqrt{t-1}}.\end{eqnarray}
Obviously, this is a contradiction.

When $g(k,q)(t-1)$ is not an integer, by setting $$t' = \frac{\lceil g(k,q)(t-1)\rceil}{g(k,q)}+1,$$ we have $t'> t$ and $g(k,q)(t'-1)$ is an integer. Utilizing the nondecreasing monotonicity of $\gamma\left(\mu\left(t,\theta\right),\theta\right)$ on $t\geq0$ for fixed $\theta$ presented in Lemma \ref{lemma4}, we can get
\begin{eqnarray}\delta_{g(k,q)(t'-1)+k}&=&\delta_{g(k,q)(t-1)+k}\nonumber\\
&<&\gamma\left(\mu\left(t,\frac{g(q,k)}{k}\right),\frac{g(q,k)}{k}\right)\nonumber\\
&<&\gamma\left(\mu\left(t',\frac{g(q,k)}{k}\right),\frac{g(q,k)}{k}\right),\nonumber\end{eqnarray}
which can be deduced to the former case. Hence we complete the proof. \qed

In order to prove the result Theorem \ref{theorem1}, we need another important concept in the RIP framework the restricted orthogonal constants (ROC) proposed in \cite{CT1}.
\begin{Definition} Suppose $\Phi\in\mathbb{R}^{m\times n}$, define the restricted orthogonal constants (ROC) of order $k_1, k_2$ as the smallest non-negative number $\theta_{k_1, k_2}$ such that
\begin{eqnarray}\label{theta} \left|\left\langle\Phi h_1,\Phi h_2\right\rangle\right|\leq\theta_{k_1, k_2}\|h_1\|_2\|h_2\|_2\end{eqnarray}
for all $k_1$-sparse vector $h_1\in\mathbb{R}^{n}$ and $k_2$-sparse vector $h_2\in\mathbb{R}^{n}$ with disjoint supports.\end{Definition}
\vspace{2mm}
\noindent\textbf{Proof of Theorem \ref{theorem1}}\\
Similar to the proof of Theorem \ref{theorem}, we only need to check for all $h \in \mathcal{N}(\Phi)\setminus\{0\}$,
$$\|h_{max(k)}\|_{q}^{q}<\|h_{-max(k)} \|_{q}^{q}.$$
 Suppose there exists $h \in \mathcal{N}(\Phi)\setminus\{0\}$, such that $\|h_{max(k)}\|_{q}^{q}\geq\|h_{-max(k)} \|_{q}^{q}$. Set $\alpha=k^{-1/q}\|h_{max(k)}\|_{q}$. From the proof of Theorem \ref{theorem}, we have $\|h_{-max(k)}\|_1\leq g(q,k)\alpha\leq\lceil g(q,k)\rceil\alpha$ and $\|h_{-max(k)}\|_\infty\leq \alpha$. Then it follows from Lemma 5.1 in \cite{CZ4} that
\begin{eqnarray}
&&\left|\left\langle\Phi h_{max(k)},\Phi h_{-max(k)}\right\rangle\right|\nonumber\\
&\leq&\theta_{k,\lceil g(q,k)\rceil}\|h_{max(k)}\|_2\sqrt{\lceil g(q,k)\rceil}\alpha\nonumber\\
&\leq&\theta_{k,k}\sqrt{\frac{\lceil g(q,k)\rceil }{k}}\|h_{max(k)}\|_2\sqrt{\lceil g(q,k)\rceil}\alpha\nonumber\\
&\leq&\theta_{k,k}\frac{\lceil g(q,k)\rceil }{k}\|h_{max(k)}\|_2^2,\nonumber\end{eqnarray}
where the first inequality holds by Lemma 5.4 in \cite{CZ4} and the second inequality by (\ref{hold}). Thus from the condition $$\delta_{k}+\theta_{k,k}\frac{\lceil g(q,k)\rceil }{k}<1,$$ it follows that
\begin{eqnarray}
0&=&\left|\left\langle\Phi h_{max(k)},\Phi h\right\rangle\right|\nonumber\\
&\geq&\left|\left\langle\Phi h_{max(k)},\Phi h_{max(k)}\right\rangle\right|-\left|\left\langle\Phi h_{max(k)},\Phi h_{-max(k)}\right\rangle\right|\nonumber\\
&\geq&(1-\delta_{k})\|h_{max(k)}\|_2^2-\theta_{k,k}\frac{\lceil g(q,k)\rceil }{k}\|h_{max(k)}\|_2^2\nonumber\\
&=&(1-\delta_{k}-\theta_{k,k}\frac{\lceil g(q,k)\rceil }{k})\|h_{max(k)}\|_2^2\nonumber\\
&>&0.\nonumber\end{eqnarray}
Obviously, this is a contradiction. By Lemma 3.1 in \cite{CZ4},
\begin{numcases}{\theta_{k,k}<}
~~~~2\delta_{k},  ~~~~~~\text{for any even}~ k\geq2,\nonumber\\
\frac{2k}{\sqrt{k^2-1}}\delta_{k}~~\text{for any odd}~k\geq3.\nonumber\end{numcases}
Hence, when $k\geq2$ is even, it yields that
$$\delta_{k}+\frac{g(q,k)}{k}\theta_{k,k}<\left(1+\frac{2\lceil g(q,k)\rceil }{k}\right)\delta_{k},$$
and when $k\geq3$ is odd, it generates that
$$\delta_{k}+\frac{g(q,k)}{k}\theta_{k,k}<\left(1+\frac{2\lceil g(q,k)\rceil}{\sqrt{k^2-1}}\right)\delta_{k}.$$
Therefore the theorem is proved. \qed

\section*{Acknowledgement}
The work was supported in part by the National Basic Research Program of China (2010CB732501), and the National Natural Science Foundation of China (11171018, 71271021).


\vskip2mm
\noindent\footnotesize{\textbf{Shenglong Zhou} is a PhD student in Department of Applied Mathematics, Beijing Jiaotong University.  He received his BS  degree
from Beijing Jiaotong University of information and computing science in 2011. His research field is theory and methods for optimization.}
\vskip2mm
\noindent\footnotesize{\textbf{Lingchen Kong}} is an associate Professor in Department of Applied
Mathematics, Beijing Jiaotong University.  He received his PhD  degree in Operations Research from Beijing Jiaotong University in 2007.
From 2007 to 2009,  he was a Post-Doctoral Fellow of Department of
Combinatorics and Optimization, Faculty of Mathematics, University
of Waterloo, Canada. His research interests are in sparse optimization, mathematics of operations
research.
\vskip2mm
\noindent\footnotesize{\textbf{Ziyan Luo}} is a lecturer in the State Key Laboratory of Rail Traffic Control and Safety, Beijing Jiaotong University. She received her PhD degree in Operations Research from Beijing Jiaotong University in 2010. From 2011 to 2012, she was a visiting scholar in Management Science and Engineering, School of Engineering, Stanford University, USA. Her research interests are in sparse optimization, semidefinite programming and interior point methods.
\vskip2mm
\noindent\footnotesize{\textbf{Naihua Xiu}} is a Professor in Department of Applied Mathematics, Beijing Jiaotong University.  He received his  PhD  degree in Operations Research from Academy Mathematics and System Science of the Chinese Academy of Science in 1997. He was a Research Fellow of City University of Hong Kong from 2000 to 2002, and he was a Visiting Scholar of University
of Waterloo from 2006 to 2007. His research interest includes variational analysis, mathematical optimization,  mathematics of operations
research.

\end{document}